\documentclass{aa}

\usepackage{graphics}

\begin{document}

   \title{Second parameter globulars and dwarf spheroidals around the
         Local Group massive galaxies: what can they evidence?}

   \author{V. Kravtsov}

   \offprints{V. Kravtsov}

   \institute{Sternberg Astronomical Institute, University Avenue 13,
              119899 Moscow\\
              \email{scorpi@sai.msu.ru}
              }

   \date{Received 29 April 2002 / Accepted 28 August 2002}

   \abstract{
   We suggest that the majority of the "young", so--called "second
   parameter" globular clusters (SPGCs) have originated in the outer
   Galactic halo due to a process other than a tidal disruption of the dwarf
   spheroidal (dSph) galaxies. Basic observational evidence regarding both
   the dSphs and the SPGCs, coupled with the latest data about a rather large
   relative number of such clusters among globulars in M33 and their low
   portion in M31, seems to be consistent with the suspected process. It might
   have taken place within the system of the most massive galaxies of the
   Local Group (LG) at the earliest stages of their formation and evolution.
   We argue that the origin and basic characteristics of the SPGCs can
   naturally be explained as a result of mass outflow from M31, during and
   due to formation of its Pop. II stars, and subsequent accretion of gas
   onto its massive companions, the Galaxy and M33. An amount of the gas
   accreted onto the Milky Way is expected to have been quite enough for the
   formation in the outer Galactic halo not only of the clusters under
   consideration but also a number of those dSph galaxies which are as young
   as the SPGCs. A less significant, but notable mass transfer from the
   starbursting protoGalaxy to the massive members of the LG might have
   occurred, too.
   }

\maketitle

\section{Introduction}

   Color--magnitude diagram (CMD) studies of the Galactic globular clusters
   (GCs) accumulated in 1960-ies have convincingly
   shown that metallicity is not the only parameter governing the
   clusters' horizontal branch (HB) morphology and that some other
   parameter(s) affect the distribution of stars on the HB. In other words,
   a "second parameter" effect takes place: at a given  metallicity,
   GCs with the second parameter phenomenon exhibit redder HB morphology.
   Thanks to numerous CMD studies, with limiting magnitude below the main
   sequence turnoff, of GCs in the Milky Way and in its nearest satellites,
   age has been revealed to play an important role in the second parameter
   effect. However, different investigators make contradictory conclusions
   regarding whether or not it is mainly responsible for the effect
   (for comparison, see Stetson et al. \cite{stetson} and Sarajedini et
   al. \cite{saraj97}; or VandenBerg \cite{van}).

   We stress the point that the second parameter problem itself is beyond
   the scope of the present paper. However, since the Galactic GCs under
   study have historically been called SPGCs, it is this term that is used
   hereafter to denote GCs younger than primeval ones, but with
   approximately the same mean metallicity, close to $[Fe/H] \sim -1.6$.

   It is now well recognized (for instance, Zinn \cite{zinn}; van den
   Bergh \cite{vdB93}; Lee et al. \cite{lee}) that GCs in the outer halo of
   the Milky Way form two subpopulations, "old" and "young". The latter
   consists of the SPGCs displaying predominantly red HB morphology.
   Presently, there is little doubt about the younger ages ($\sim2$ Gyr) of
   the SPGCs as compared to the old GCs in the Galactic halo (e.g., Lee et al.
   \cite{lee}). In addition to such age characteristics, their predominant
   location in the halo at Galactocetric distance ${R_{GC}}>8$ kpc poses the
   fundamental problem of the origin of these clusters. It is a
   long--standing, poorly understood
   problem that arose more than three decades ago. Revealing the cause
   of these GCs in the outer Galactic halo would be beneficial
   in understanding their formation and the early history of the Milky
   Way, and other related questions.

   By proceeding from the tentative inference that age
   is responsible for the second parameter, Searle \& Zinn (\cite{sz}), in
   their scenario of halo formation, hypothesized that the origin of
   SPGCs was in gaseous condensations, "transient protogalactic fragments
   that continued to fall into dynamical equilibrium with the Galaxy for
   some time after the collapse of its central regions had been completed".
   In this scenario the authors also considered the possibility of merging
   of low--mass galaxies with the Milky Way. Later, owning to development of
   the cold dark matter theories of galaxy formation, some investigators
   tentatively identified these hypothetical transient fragments with "dark
   halos" which, in turn, were found in the dSph galaxies, real
   objects with both the appropriate masses of visible matter and signs of
   a rather high dark matter content (see, among others, the review of
   various topics related to dSphs by Gallagher \& Wyse \cite{galwys};
   review by Mateo \cite{mateo} ). Moreover, a number of  important
   observations -- in particular, the discovery by Ibata et al. (\cite{ibata})
   of the Sagittarius dSph that shows obvious signs
   of tidal disruption; findings about the second parameter effect
   displayed by the old stellar populations in dSphs and by GCs
   belonging to them (e.g., among others, Buonanno et al. \cite{buonan85};
   Suntzeff et al. \cite{suntz};  Demers \& Irwin \cite{demirw}  );
   a highly ordered spatial distribution of the Galactic dSph galaxies, as
   noted by Kunkel \& Demers (\cite{kundem}) and Lynden-Bell
   (\cite{lynden}), along with a close resemblance between three--dimentional
   distribution of SPGCs having the reddest HB morphology and some of the
   dSphs (Majewski \cite{majew}) -- are evidence suggesting that
   the origin of the Galactic SPGCs is a
   consequence of tidal disruption/stripping (TD/S), during a Hubble time,
   of the formerly more numerous dSph--like systems or larger Galactic
   companion(s). Thus, the role of such a mechanism is at present
   widely believed to be crucial, and its reliability with regard to
   the formation of the Galactic halo is currently being investigated
   (for instance, Morrison et al. \cite{morris}; Harding et al. \cite{hard}).

   Explaining the population of Galactic SPGCs by TD/S of the respective
   ancestral galaxies, dSph or some other satellites to the Milky Way
   appears to be a natural and well grounded approach to the problem.
   However, {\it even if it was an absolute fact it
   would simply replace the problem by other one(s) related to the origin of
   the ancestors themselves}. Indeed, the Galactic SPGCs are, on the one hand,
   younger than the old halo globulars, but, on the other hand, they
   should have been among the oldest constituents of their
   hypothetical parent (dSph) systems. Therefore, we have to
   conclude that these tidally dissolved dSphs are younger (by $\sim2$
   Gyr or so) than the oldest objects populating the Galactic spheroid.
   {\it Then, we need to explain the mysterious emergence of the "young"
   satellite galaxies in the outer Galactic halo, in about two Gyr
   after the beginning of its formation, as well as their
   relatively rapid subsequent dissolution by the Galactic tidal
   forces.} Moreover, different facts and their detailed analysis (see
   Section 2) can militate against such an attractive "destructive"
   mechanism as being the major one responsible for the Galactic SPGC
   system formation.
   
   Instead of (or in addition to) this we have recently suggested (Kravtsov
   \cite{krav}), and here we analyze in detail, a "creative" process
   which would perhaps be able to explain the origin of populations of both
   SPGCs and dSphs, belonging not only to the Galaxy but also to two other
   massive galaxies of the LG. Specifically, M31 includes six dSph systems
   (Grebel \cite{grebela}), while M33 has no dwarf companions, but according
   to data of Sarajedini et al. (\cite{saraj98}) it has appreciable population
   of SPGCs.

   The role of the dSph--consuming process in the formation of the Galactic
   SPGCs population and populations of sush GCs in M33 and M31 is analyzed
   in the next section, taking into consideration diverse observational data
   obtained to date. In Section 3, we summarize the basic
   characteristics of both the population of Galactic SPGCs and the dSph
   galaxies, which may be consistent with and explained by the processes
   described in Section 4.

\section{Is dSph--consuming process responsible for the origin of the SPGCs?}

To achieve a more adequate understanding of the nature of the Galactic SPGC
population, observational information about such clusters in other
massive galaxies of the LG is very valuable. Preliminary relevant results
obtained to date seem to increase doubts about the TD/S of dSphs as
{\it the universal mechanism of and main contributor to}
the formation of populations of SPGCs in the three most massive galaxies of
the LG. Contrary to expectations following from the large mass of M31 and
hence its comparatively high destructive ability, among GCs studied in the
galaxy there are only hints of the existence of SPGCs (Rich et al.
\cite{rich}). This means that the relative number of these clusters among
GCs in M31 is presumably low. Selection effects can take place though, taking
into account that among Galactic globulars the SPGCs have, on average, lower
concentrations and luminosities (van den Bergh \cite{vdB96}). At the same
time, a surprisingly large number of GCs, eight out of ten in the sample
studied by Sarajedini et al. (\cite{saraj98}) in M33, have turned out to
show exclusively red HB morphology at the mean metallicity $[Fe/H]=-1.6$,
i.e. they are the SPGCs. The authors carefully selected most probable
candidates for the halo objects of M33, and hence relative number of SPGCs
in the halo of the Triangulum nebula is apparently high. This is further
supported by the latest investigation by Chandar et al. (\cite{chand}). They
found that $\sim 60\%$ of 17 kinematically selected halo clusters of M33
appear to be "young" ones, "a much higher fraction of young to old halo
clusters than found in the Milky Way". The observed clusters' projected
distances from the galaxy center, at the given mass of the galaxy, do not
exclude the origin of SPGCs in M33 due to tidal disruption of hypothetical
dSphs (but not to stripping, because we see no companions to M33). However,
the observations by Sarajedini et al. and by Chandar et al. (\cite{chand})
coupled with recent result of Chandar et al. (\cite{chandar}) are able to put
significant constraints on such a possibility. The latter authors have
estimated the total number of GCs in M33 to be $75\pm14$, which gives
a specific frequency significantly higher than in other late--type spirals.
It is most probable that this surprising excess is due to the presence of
SPGCs. Therefore, in the
framework of the dSph--consuming view of the origin of SPGCs we should
again assume, as in the case of our Galaxy, the mysterious emergence of
the hypothetical "young" satellite dSph galaxies in the halo of M33
a few Gyr after the beginning of its formation. Moreover, it is necessary
to accept a surprisingly complete disruption of all these dwarfs, as well as
their preferably high mass, since only the high--mass dwarfs would be able
to form GCs. These processes and characteristics seem to be fairly strange and
artificial, and they could hardly be real in such a combination as applied to
M33. But, there is little doubt that some specific process(es) occurred in
the halo of M33 that led to both the prolonged formation of GCs and their
obvious surplus population in the galaxy compared to those in many other
late--type spiral galaxies.

Important constraints on the Galactic halo formation by
the dSph--consuming process come from the latest abundance investigations.
Shetrone et al. (\cite{shetrone}) have determined and analyzed abundance
ratios for more than 20 elements in spectra, obtained with the Keck I
telescope, of red giant stars belonging to the Draco, Sextans, and Ursa
Minor dSphs. From a comparison of the measured values with published ones
for the Galactic halo stars, a number of essential differences between them
was found, particularly in $[\alpha/Fe]$. Hence, the authors have
concluded that their observations do not support the notion "that the
Galactic halo has been assembled {\it entirely} through the disruption of
very low--luminosity dSph galaxies like the three galaxies" studied by them.
From a similar comparison of element abundance ratios
measured for the Galactic halo stars and dSph stars within the same range of
metallicity, Fulbright (\cite{fulb}) has made even stronger claims that
the former stars in the solar neighborhood are unlikely to have been
constituents of disrupted dSphs similar to the studied ones.

Fairly strong limits to the Galactic halo formation via dSph satellite
merging have been set by Unavane et al. (\cite{unavane}) and Gilmore et al.
(\cite{gilmore}) as well. According to their estimates no more than
some $10\%$ of the halo could have formed due to this process over
approximately the last 10 Gyr. The latter authors also conclude that "either
the dwarfs merged before they formed stars, or they never formed".

Interesting supplementary constraints on the Galactic SPGCs were analyzed by
van den Bergh (\cite{vdB96}).

In addition to the above--mentioned, indirect studies of possible
TD/S of the hypothetical dwarf galaxies in the past, direct investigations
were carried out regarding the impact of the TD/S on mass loss from the
dSph "survivors" really existing in the outer Galactic halo. Unfortunately,
the respective results obtained to date for some dSph galaxies are very
different, and they now cannot provide a decisive answer to the problem.
Apart from the conventional process of the TD/S of the Sagittarius dSph
galaxy, indications of an extratidal population around the Ursa Minor dSph
are reported by Martinez-Delgado et al. (\cite{mart}). A substantial
extratidal population from the Carina dSph galaxy
has been found by Majewski et al. (\cite{majew00}), who have determined the
rate of current mass loss from the galaxy to be surprisingly high, namely of
order of $27\%$ of its mass every Gyr. Hence, they evaluate, under some
assumptions, that 14 Gyr ago the Carina dSph was 100 times more massive than
now. At the same time, Odenkirchen et al. (\cite{oden}) and Aparicio et
al. (\cite{aparicio}) have found no tidal tails or an extratidal
extension of the stellar populations belonging to the Draco dSph galaxy.
For comparison with some of the above findings it is interesting to note the
latest results obtained by Hayashi et al. (\cite{hayashi}), modeling the
dark matter halos surrounding dSph satellites of the Milky Way. They find
that these halos have circular velocity curves with peaks at essentially
higher velocities than determined from the stellar velocity dispersion, and
that the tidal radii of the dSphs are larger than those derived from
observations.

Taken at face value, these observations may imply that a role played by the
dSph--consuming process in the formation of the population of SPGCs in the
outer Galaxy, as well as populations of these clusters around the LG massive
spirals, is secondary. For this reason, we suggest other process as being
important. In our approach we proceed from the recognized starting point that
there is an apparent relationship between the dSphs and SPGCs. However,
{\it we assume that both kinds of objects originated from the same cause
rather than the dSph--consuming process resulted in the populations of
SPGCs.}

\section{Summary of the basic characteristics}

   The relationship between the Galactic populations of dSphs and SPGCs is
   reliably shown by their basic characteristics which, in turn, should
   be consistent with the process suspected to be responsible for the origin
   of these objects. These characteristics are summarized below.

   \noindent {\bf --}  {The age of the SPGCs, as mentioned above, differs
   from that of the old GCs by a few Gyr, typically $\sim2$ Gyr (but see
   age data for some "young" globulars in VandenBerg \cite{van}).}

   \noindent {\bf --}  {These clusters exhibit peculiar kinematic
   characrestics. They have, on average, marginally retrograde or zero
   rotation, whereas old GCs show direct motion with lower dispersion.
   According to Da Costa (\cite{dacosta}) the mean velocities of rotation are
   ${V_{rot}}=-45\pm81$ km s$^{-1}$ and ${V_{rot}}=58\pm24$ km s$^{-1}$,
   respectively.}

   \noindent {\bf --}  {Metallicities of the SPGCs, as noted by Rodgers
   \& Paltoglou (\cite{rodpal}) and confirmed by van den Bergh (\cite{vdB93}),
   fall in a quite narrow range, within $\Delta [Fe/H]=0.4$ dex,
   with no relation between metallicity and Galactocentric distance.
   However, Yoon \& Lee (\cite{yoonlee}) argue that the majority of the most
   metal--poor ($[Fe/H]<-2.0$) Galactic GCs are younger, by $\sim1$ Gyr,
   than the oldest globulars, and find that these clusters "display a planar
   alignment in the outer halo", close to the Magellanic Plane.}

   \noindent {\bf --}  {Majewski (\cite{majew}) has shown that
   three--dimentional distribution of the SPGCs with reddest HB morphology is
   such that they appear to populate the so-called "Fornax--Leo--Sculptor
   stream" consisting of five dSph galaxies orbiting the Milky Way. He
   has argued that all these objects "may be related through origin in a
   common Galactic accretion event".}

   \noindent {\bf --}  {Moreover, Galactic dSphs themselves show ordered
   distribution on the sky, with the plane formed by them being nearly
   orthogonal
   to the Galactic plane (Kunkel \& Demers \cite{kundem}; Lynden-Bell
   \cite{lynden}; Majewski \cite{majew}). A similar ordered distribution is
   also displayed by dSphs belonging to the Andromeda galaxy. Both dSph
   populations around the two parent galaxies apear to be distributed in
   space either roughly near the same plane or near two planes with a low
   inclination between them, as may be seen, for example, in the results
   of Hartwick (\cite{hartwick}) or in the Figures 1 and 3 from Grebel
   (\cite{grebelb}) and Mateo (\cite{mateo}), respectively.}          

   \noindent {\bf --}  {In contrast to dwarf irregular galaxies, which are
   more or less uniformly distributed across the LG, dSphs show
   obvious concentration to the Galaxy and M31 (e.g.,Einasto et al.
   \cite{einasto}; van den Bergh \cite{vdB94}; Grebel \cite{grebela}).}

   \noindent {\bf --}  {CMDs of dSphs and their GCs, like those of the SPGCs,
   are strongly affected by the second parameter. This may be evidence that
   ages of at least some of the objects are younger than the old globulars.
   For instance, according to Suntzeff et al. (\cite{suntz}) the Sextans dSph
   galaxy "may be a few Gyr younger than the typical Galactic globular
   cluster".
   Saragedini et al. (\cite{saraj02}) conclude that the Cetus dSph is 2--3
   Gyr younger than Galactic GCs with the same metallicity, provided that age
   is primarily responsible for the second parameter. On the contrary,
   Buonanno et al. (\cite{buonan98}) find no significant difference
   in age between GCs belonging to the Fornax dwarf galaxy and Galactic GCs
   M68 and M92. Also, the Ursa Minor dSph galaxy is estimated to be as
   old as the latter globular (Mighell \& Burke \cite{migbur}). It should
   be noted here that, due to a number of well known reasons, the accuracy
   of the relative age estimates for the Galactic dSph galaxies is hardly
   better than 1 Gyr, even at best.
   In any case, it is quite probable that at least a number of the presently
   observed dSphs belonging to the Milky Way are not the primeval
   objects in the sense that their oldest stellar populations are younger
   than the old Galactic globulars. {\it One can suggest that either such
   dSphs were formed in relation to some secondary process(es) or initial
   starbursts in the protodSphs ("dark halos") were delayed until they were
   triggered by these processes.}}

   \noindent {\bf --}  {Note also that both the number of dSphs orbiting
   the Galaxy and the total mass of luminous matter that they comprise, as
   one can deduce from the compilative data of Grebel \cite{grebela}, are
   apparently larger when compared to dSph systems belonging to M31.}

\section{Putative consequence of the mass outflows}

It has been argued since 1970s (e.g., Larson \cite{larson}; Bookbinder et al.
\cite{book}; Marochnik \& Suchkov \cite{marsuch}; Berman \& Suchkov
\cite{bersuch}) that some properties of
galaxies, particularly chemical ones, and those of hot gas in clusters of
galaxies imply a considerable mass loss which may have occurred in the
early epoch of formation and evolution of the galaxies. This is mainly due
to the mass outflows caused by supernovae--driven winds during the most
powerfull (initial) star formation event(s). Such a loss of gas by galaxies
such as M31 and the Milky Way has been evaluated to amount up to a half mass
of the respective protogalaxy, i.e. $\Delta M \sim10^{11} M_{ \odot}$.

These conclusions, reached in earlier studies, seem to be in general
agreement with the latest results achieved on the basis of more advanced
observational and theoretical investigations of starbursting galaxies at
different redshift. The high--redshift Lyman break galaxies (LBGs) are
suggested to be progenitors of the present--day massive spheroids being, in
particular, components of luminous early--type spirals (Friaca \& Terlevich
\cite{friter}, for instance). Starburst--driven galactic superwinds observed
in LBGs play important role in the mass outflow from them and in enrichment
of the intergalactic medium. As summarized by Heckman (\cite{heckman}, and
references therein), "the outflows carry mass out of the starburst at a rate
comparable to the star--formation rate". This implies that outflows from a
starforming protogalaxy, like M31, during the most powerful episode(s) of its
Population II formation, that is (are) assumed to last $\sim2$ Gyr with the
star formation rate $\sim100 M_{\odot}$ yr$^{-1}$ (Granato et al.
\cite{granato}) would be able to carry off the protogalaxy a total mass of
gas somewhat exceeding
$\sim10^{11} M_{ \odot}$. Because according to Pettini et al. (\cite{pet})
typical speeds of such a gas in LBGs range from a few $10^{2}$ to $10^{3}$
km s$^{-1}$, some tens of percent of this mass, i.e. up to $\Delta M
\sim10^{11} M_{\odot}$ (that is in agreement with the former estimate), with
velocities exceeding the escape velocity of M31,
could have left the galaxy.

If so, then portion of the gas that escaped from M31
could be accreted onto its massive companion(s), our Galaxy (and M33), as it
might be expected from the well--known processes taking place in binary star
systems. We can easily estimate an order of magnitude of a time delay
between starting an
expansion of the gas from M31 and its subsequent accretion onto the Galaxy. If
we build upon both the present distance between the two galaxies ($\sim0.7$
Mpc) and the ratio of their mass (${M_G}/{M_A} \sim1/1.5$) then we find that
the time spent by the gas to reach a region of the Lagrangian point between
the two galaxies ($\sim0.4$ Mpc) may be as long as 1 Gyr at a mean
velocity of the expanding gas of the order of $\sim400$ km s$^{-1}$. The
assumed duration ($\sim2$ Gyr) of mass loss from M31 combined with the above
time delay can explain, in particular, the mean age difference between the
SPGCs and primeval globulars in our Galaxy.

Yet one key quantity is an upper limit of mass of the gaseous material $\delta
M$ that might have been accreted onto the Galaxy if the total amount of
gas lost by M31 was $\Delta M (\sim10^{11} M_{\odot})$. For this purpose we
have applied a formula deduced by Postnov (\cite{postnov}) in the approach
of gas accretion in gravity potential $\varphi \sim{{1}\over{R}}$ of the
field of the less massive companion in a binary system. By substituting the
corresponding values in the formula, we obtain: 

$$\delta M \leq{{1}\over{4}}{ \bigg({{M_G} \over{M_A}} \bigg)}^2\Delta M
\sim{{1}\over{10}}\Delta M \sim10^{10} M_{ \odot}$$

Even if a mass of really accreted gas was an order of magnitude lower than
the upper limit estimated and if, in turn, approximately ten percent of
this mass was converted into objects, i.e., $\sim10^{8} M_{\odot}$,
that would be equal to a total mass of the bulk of SPGCs, a few
dSph galaxies, and also of field stars with their total mass 1 -- 1.5 order
of magnitude higher compared to the mass of all the SPGCs, which might have
formed due to such a process in the outer halo of our Galaxy. The typical
cluster and dSph masses are accepted to be $\sim10^{5} M_{\odot}$ and
$\sim10^{7} M_{\odot}$, respectively. It is worth of noting that from a
detailed analysis of the space motions of the Galactic halo
stars, Chiba \& Beers (\cite{chibeers}) find both an excess number of
high--eccentricity stars and a sharp discontinuity, near zero, of
the mean rotational velocities to occur near the same metallicity,
$[Fe/H] \sim -1.7$. They suggest that a significant fraction of stars with
metallicity near this value might form from infalling gas with the
corresponding chemical abundance during the early formation of the Galaxy.
Intriguingly enough, average metallicities of SPGCs in the Galaxy and
M33, falling near $[Fe/H] \sim -1.6$, are very close to the above value.

Similarly, by applying the above formula to the case of M33 and taking into
account that its mass is a factor of ten lower than the mass of M31, it is
possible to conclude that objects assumed to form from a gas accreted onto
M33 might include some dozen (or more) GCs and a population of field stars
too.

Apart from M31, the early Galaxy is also expected to lose a sizable
amount of partially enriched gas, though its overall mass should apparently
be less than in the case of the former galaxy. Therefore, a less significant
mass transfer, from the star--forming protoGalaxy to its massive companions,
along with the respective consequences are suspected to have taken place too.
Here, it is worth of mentioning that, by analyzing potometric properties of
the paired early--type galaxies, Demin (\cite{demin}) assumed the
possibility of mass--exchange processes between such galaxies
at the early stage of their evolution, and Demin \& Sazhin (\cite{demsazha},
\cite{demsazhb}) studied this possibility. In particular, in the latter
paper the authors conclude that mass exchange between companions in a binary
galactic system can be more preferable than a mass loss from the system.

Of course, we here use simple evaluations which are preliminary. However,
they appear to be sufficient to show, in principle, the expected scale
and consequences of mass transfer between the most massive members of the LG.
Additionally, the numerous observations of present--day formation of
massive star clusters and dwarf galaxies in the tidal tails of
interacting galaxies (e.g., Gallagher et al. \cite{galetal} and references
therein) may be important indirect observational evidence supporting
the formation of such objects due to the suggested type
of interaction between the LG massive galaxies during formation of their
spheroids. At this stage of their formation, a high portion of mass of each
of the starbursting galaxy remains in the form of a gas, part of which is
contracting to the central regions of the spheroids, and another part is
expanding out of them. It is gas in the outer regions of the star--forming
protoGalaxy that could primarily collide with an accelerated infalling
stream of gas which was being transferred from the Andromeda protogalaxy
through a region of the Lagrangian point between the early galaxies.
Supersonic gas motion and accretion, under a speed of the gaseous material
relative motion reaching at least a few $10^{2}$ km s$^{-1}$, should have
led to shockwaves. Indeed,
sound speed in gas with $T \approx 10^{6} K$ is near 150 km s$^{-1}$,
whereas the speed of the accelerated infalling gas might be not less
than $200\div300$ km s$^{-1}$. Shockwaves are known to be powerful factor
triggering and essentially increasing efficiency of star formation. Perhaps,
the conditions favourable for the formation of stars and clusters were
more or less met at a distance of several tens of kiloparsecs from the
Galactic centre. The efficiency of star formation would probably be highest,
irrespective of Galactocetric distance, if it happened that the infalling
stream collided with relatively dense gas contained in the "dark halos".
It is important to empasize that, by analogy with a mass transfer between
companions in an interacting binary star system, the passage of gas through
the Lagrangian point between spheroids of the two protogalaxies should have
led the infalling gas to become denser and to have more or less ordered
distribution and motion around the protoGalaxy, close to the plane defined by
the relative motion of M31 and the Milky Way at that moment. Therefore, the
objects that could
form from the infalling gas, thanks to the mass transfer from M31 to the
Galaxy, would exhibit signs of the respective kinematics and spatial
distribution around the latter galaxy in its outer parts. The same should be
correct for gas accretion onto M33. The observed
characteristics of the aforementioned stars with metallicity around $[Fe/H]
\sim-1.7$  and populations of both the Galactic SPGCs and dSphs seem to
agree well with the objects' origin due to such a process.
Kinematic and spatial characteristics of SPGCs in M33 also suggest their
accretive origin in the halo of the galaxy (Chandar et al. \cite{chand}).

Similarly, it is not excluded that the infalling gas might have collided
with the interstellar matter in the outer parts of the early LMC, the
largest Galactic satellite. This could result in the induced formation of
additional globulars more weakly bound to the LMC, which were subsequently
captured by the Galaxy.

Accepting the proposed and discussed interaction between the Milky Way, the
Andromeda galaxy, and the Triangulum nebula in their early history, we can
naturally explain (in the present section) the existence and the basic
characteristics (Section 3) of SPGCs and (at least) some dSphs orbiting
these galaxies. We omit the problem dealing with the relationship between
the number of SPGCs and field stars formed in the same star--forming
process. It is related to the more general, complex problem of GC (massive
cluster) formation and its efficiency (e.g., Larsen \& Richtler
\cite{larich}; Elmegreen \cite{elmeg}). Note only that for the populations of
"normal", metal--poor and metal--rich GCs belonging to either elliptical
galaxies or spheroids of spirals such a relationship may vary essentially
(e.g., Harris et al. \cite{harris} and Durrell et al. \cite{durrell}).

Observations of high redshift structures of galaxies, such as reported by
Venemans et al. (\cite{venem}), would perhaps be able to shed light on the
problem of the mass transfer between massive galaxies at early stage of
their evolution.

\section{Conclusions}

   The present paper discusses long--standing problem of
   the origin of the Galactic SPGCs.
   We show that there arise questions about the currently popular and
   conventional view on the origin of a population of these globulars as a
   result of TD/S of dSph galaxies, especially taking into account
   recent relevant observational data including those on the populations of
   such globulars in the two other most massive galaxies of the LG, M31 and
   M33. On the one hand, answering these questions requires a number
   of a priori assumptions, some of which seem to be rather artificial and
   contradictory, about hypothetical events and objects. On the other hand,
   we argue that even if this view is correct, the problem is merely
   replaced by other one(s) related to the origin of dSphs themselves.
   In order to interpret and reconcile the various observational data on
   SPGCs (and probably dSphs) belonging to the three spirals of the LG we
   propose here other scenario. It is based on the realistic suggestion about
   natural consequences of mass loss from M31 in its early history,
   namely, subsequent partial accretion of the gaseous material onto the
   early Galaxy and M33 should have contributed to the formation of their
   (outer) halos.

   In addition, we draw attention
   to the possible interaction between massive galaxies at the early stage of
   their formation and evolution in pairs, groups, and perhaps in small
   clusters. In contrast to the relatively well--known and recognized
   consequences of a tidal interaction between galaxies, the suspected role
   of the interaction we deal with seems to be presently missing or
   underestimated. We suggest that if some starbursting massive protogalaxy,
   like M31 or our Galaxy, really loses a substantial amount of its gas, then
   the partially enriched gas expanding out and escaping from the protogalaxy
   not only contributes to the intergalactic medium but also should partially
   be trasferred to massive companion(s) of the protogalaxy. Because of the
   sizable amount of the gaseous material (probably, around a few per cent of
   mass lost by the starbursting massive protogalaxy) trasferred to and
   accreted onto the massive companions, they may experience additional
   large--scale starbursts leading to the formation, out of this material, of
   supplementary objects (stars, star clusters, and even dwarf galaxies)
   with particular appropriate characteristics.
   Therefore, strictly speaking, such galaxies do not form and evolve
   as completely isolated entities, and they may contain populations of
   SPGCs (and probably dSphs) as tracers of the interaction under study.
   
   We suppose that just this kind of interaction occurred between the most
   massive galaxies of the LG and lasted approximately 2 -- 3 Gyr after the
   beginning epoch of their formation, and it could primarily be responsible
   for the origin of both their SPGC populations and some of the dSph
   galaxies, as well as for the basic observables of these objects.

\begin{acknowledgements}
   The author is very grateful to A.A. Suchkov for his first encouraging
   comments on the idea of the scenario presented, and to
   S. van den Bergh whose attention and interest in the scenario outline
   presented in astro--ph was important moral support.
   Thanks are due to Yu.N. Efremov, K. Postnov, A.S. Rastorguev, M.V. Sazhin,
   N.I. Shakura,  Yu.A. Shchekinov, and V.V. Slavutinsky for useful
   discussions. L. Murray's assistance in improving the English of
   the paper is acknowledged. The author thanks an anonymous referee whose
   useful criticism stimulated improvement of the paper.

\end{acknowledgements}

\end{document}